\documentclass[%
 aip, amsmath,amssymb,mathrsfs, bbold, reprint]{revtex4-1}

\usepackage{graphicx}
\usepackage{dcolumn}
\usepackage{bm, bbold}

\usepackage[utf8]{inputenc}
\usepackage[T1]{fontenc}
\usepackage{mathptmx, mathrsfs}
\usepackage{etoolbox}

\makeatletter
\def\@email#1#2{%
 \endgroup
 \patchcmd{\titleblock@produce}
  {\frontmatter@RRAPformat}
  {\frontmatter@RRAPformat{\produce@RRAP{*#1\href{mailto:#2}{#2}}}\frontmatter@RRAPformat}
  {}{}
}%
\makeatother

\begin{document}
\author{Nicholas J. Harmon}
\email{harmon.nicholas@gmail.com} 
\affiliation{Department of Physics and Engineering Science, Coastal Carolina University, Conway, SC
29526, USA}
\affiliation{Department of Physics, University of Evansville, Evansville, Indiana
47722, USA}
\author{James P. Ashton}\thanks{Current address: Keysight Technologies Inc., Santa Rosa, CA 95403}
\affiliation{Department of Engineering Science and Mechanics, Pennsylvania State University, University Park, PA, 16802, USA}
\author{Patrick M. Lenahan}
\affiliation{Department of Engineering Science and Mechanics, Pennsylvania State University, University Park, PA, 16802, USA}
\author{Michael E. Flatt\'e}
\affiliation{Department of Physics and Astronomy, University of Iowa, Iowa City, Iowa
52242, USA}
\affiliation{Department of Applied Physics, Eindhoven University of Technology, Eindhoven, The Netherlands}
\date{\today}
\title{Spin-Dependent Capture Mechanism for Magnetic Field Effects on Interface Recombination Current in Semiconductor Devices}
\begin{abstract}
Electrically detected magnetic resonance (EDMR) and near-zero field magnetoresistance  (NZFMR) are techniques that probe defect states at dielectric interfaces critical for metal-oxide-semiconductor (MOS) electronic devices such as the Si/SiO$_2$ MOS field effect transistor (MOSFET). A comprehensive theory, adapted from the trap-assisted recombination theory of Shockley,  Read, and Hall, is introduced to include the spin-dependent recombination effects that provide the mechanism for magnetic field sensitivity. 
\end{abstract}
\maketitle	

Magnetic resonance experiments have given access to the microscopic details of defects inside various materials, including semiconductors and insulators, through electron spin resonance (ESR) techniques and primarily the technique referred to as electron paramagnetic resonance (EPR).\cite{Wertz1970} Increased sensitivity is provided by the closely associated ESR method called electrically detected magnetic resonance (EDMR).\cite{Chen2003, Boehme2005} This method takes advantage of spin-selection rules. Although more limited in use, as EDMR requires  current-generating transitions between spin pairs, detection occurs through sensitive electrical measurements which can be at least ten million times more sensitive than that of conventional EPR.\cite{McCamey2006}  
Magnetic resonance techniques have been particularly useful in discovering and determining properties of deep-level paramagnetic defects in semiconductors and insulators that are widely used in contemporary and prospective integrated circuits.\cite{Lenahan1998, Watkins1999} 
The electronic states lying near the middle of the band gap contribute substantially to recombination current which allows them to be accessed efficiently using EDMR. 
For instance, the P$_b$ defect, a dangling bond, appears at the interface of Si and SiO$_2$ in Si/SiO$_2$ MOSFETs;\cite{Nishi1971, Nishi1972} these centers capture charge carriers, shifting the threshold voltage, and reduce effective transistor channel mobilities.\cite{Lenahan1982, Lenahan1984, Kim1988, Vranch1988, Miki1988, Awazu1993} If these defects are present in significant numbers, which occurs when devices are irradiated or stressed by other means, they substantially limit transistor performance.\cite{Ashton2019, Fleetwood2009}
These resonant methods require a radio frequency or microwave frequency field, $B_1$, to induce spin transitions.

Recently  recombination current measurements have displayed magnetoresistive effects near zero field comparable to EDMR, whether or not an alternating field was applied\cite{Cochrane2012, Cochrane2013, Ashton2019}. The likely connection between this near-zero field magnetic resistance (NZFMR) and the same defects that play a role in EDMR suggests that NZFMR spectroscopy could be a new tool to study defects in materials such as semiconductors and insulators \cite{Harmon2020}, especially when shielded by metal contacts such as deep within integrated circuits.
For spin $S = 1/2$ defects, EDMR and NZFMR both operate by singlet-triplet transitions taking place which open or close the trap recombination channel to carrier spins. EDMR does so by changing the total spin of the pair when on resonance. When spin transitions are induced by hyperfine coupling, which  we consider here, NZFMR changes  the spin of the spin pair by transferring spin into the nuclear spin bath. In either case the change in spin allows for the recombination bottleneck to be lifted. 

In this Letter, we provide a comprehensive calculation to determine the recombination current responses, which requires modifying the conventional Shockley Read Hall description\cite{Shockley1952, Hall1952} of trap-assisted recombination, that arise from both EDMR and NZFMR. Since these effects are observed as a magnetic field, $B_0$, is swept, we broadly refer to these responses as magnetic field effects. 
 \begin{figure}[ptbh]
 \begin{centering}
        \includegraphics[scale = 0.25,trim = 20 170 0 0, angle = -0,clip]{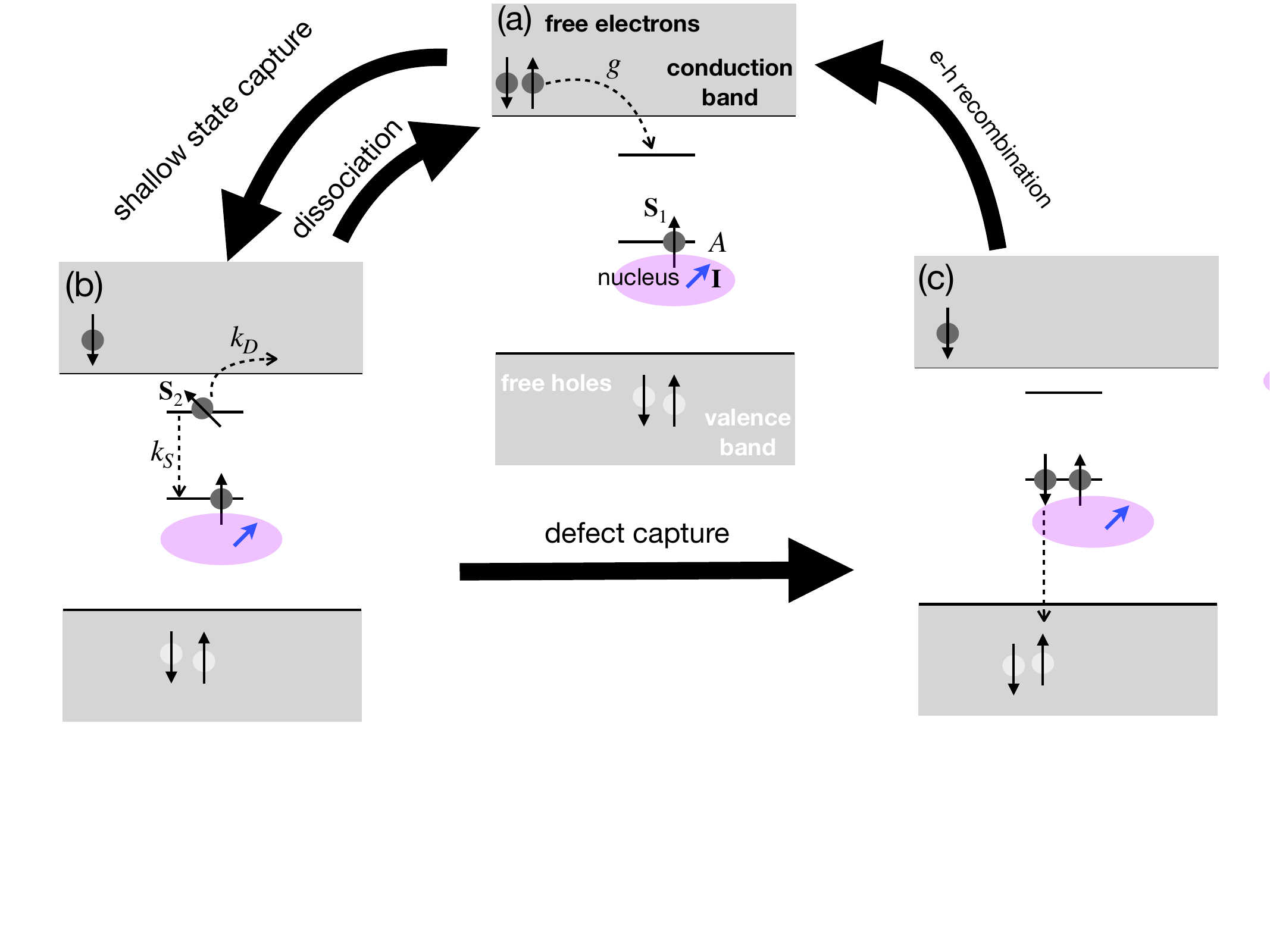}
        \caption[]
{Spin-dependent recombination through a deep-level paramagnetic defect.  (a) Free electrons encounter a defect through a shallow state. (b) The weakly localized electron can dissociate at rate $k_D$ (return to (a)). Only a singlet state can exist at the defect site in which case the shallow electron may be captured by the defect at rate $k_S$ (proceed to (c)). (c) The extra electron at the defect recombines rapidly with a free hole which  returns the defect to its paramagnetic state (a). The blue arrow represents a nuclear spin in the vicinity of the deep electron. $A$ is the hyperfine coupling constant for the defect.}\label{fig:RongModel} 
        \end{centering}
\end{figure}
Our approach utilizes a set of equations known as the stochastic Liouville equations which naturally account for the spin-selective processes. We determine line shapes of NZFMR and EDMR when a nuclear spin is represented quantum mechanically. By doing so, we are able to obtain understanding of the physics of spin-dependent recombination through deep-level paramagnetic defects. The results from our modeling suggest that NZFMR spectroscopy may serve as a new diagnostic for defects in semiconductors and insulators (especially at the interfaces between them).

Shockley, Read, and Hall \cite{Shockley1952, Hall1952} postulated a non-radiative recombination mechanism of conduction electrons and holes via traps. 
In their model, electron-hole pair recombination most effectively takes place at deep level defect centers near mid-bandgap. Thus, in measurements of interface recombination utilizing a gated diode measurement, the recombination defect centers have energy levels very near the middle of the bandgap. Recombination current results from capture of both types of charge carriers at a deep level. Consider the simple case of an electron traveling through the conduction band. When the electron encounters a deep level defect, it may fall into the deep level granted that spin selection rules are obeyed. Once electron capture takes place, the electron is available for recombination with a hole in the valence band. This process may also take place via hole capture and subsequent recombination with a conduction band electron. In the gated diode measurement, the recombination current measured through the body contact of the MOSFET has a peak which corresponds to the situation where both electron and hole densities are equal (recombination is maximized). At a given forward bias, tuning the gate voltage  achieves a maximum recombination current expressed as
\begin{equation}\label{eq:Imax0}
I_{max} = \frac{1}{2} q^2 n_i \sigma v_{th} A D_{it} |V_F| e^{q|V_F|/2 kT}
\end{equation}
where $\sigma$ is the equal electron and hole capture cross sections by the defect, $v_{th}$ is the thermal velocity of the carriers, $n_i$ is the intrinsic carrier concentration, $D_{it}$ is the areal concentration of interface traps per energy, $A$ is the gate area, $q$ is the electronic charge, $V_F$ is the forward bias voltage, $k$ is Boltzmann's constant, and $T$ is the absolute temperature. This $I_{max}$, and another when cross sections are unequal, are derived independent of spin.\cite{Fitzgerald1968}

Early on a magnetic field dependence was noted but initial models,  suggesting a spin dependence due to spin polarization, were unsuccessful in explaining magnetoresistances \cite{Lepine1972a, Lepine1972b, Lvov1977, Wosinski1977, Mendz1980}.
Kaplan, Solomon, and Mott (KSM) posited that electron and hole spins pass through an intermediary phase (which might be an exciton or a donor-acceptor pair) wherein the two spins, for some amount of time, are exclusive to one another \cite{Kaplan1978}. The pair has the option to either recombine or dissociate. Only the recombination process is spin-dependent. 
Haberkorn and Dietz\cite{Haberkorn1980} provided a more general theoretical framework to the KSM model by expressing the charge and spin dynamics through a stochastic Liouville equation. 
Rong \emph{et al.} called into question the KSM assumption that it is electrons and holes that make up the exclusive spin pair; Rong \emph{et al.} proposed intermediate states to be either an excited state of the defect or a shallow donor state near the conduction band.\cite{Rong1991} The carrier is first trapped into the higher energy localized state where it then has some probability of either dissociating back into an itinerant state or falling down into the ground state of the defect.
It is this last process which is spin-dependent due to the Pauli exclusion principle since the defect is initially paramagnetic. The singlet ground state of the charged defect may capture an electron that recombines with a hole and the process repeats with the defect becoming paramagnetic once more.\cite{Spaeth2003} These processes are depicted in Figure \ref{fig:RongModel}.
With this history in mind, Eq.~(\ref{eq:Imax0}) is insufficient for a few reasons: the electron and hole cross sections, $\sigma_e$ and $\sigma_h$, are unequal\cite{Garrett1956}; the recombination is mediated by an intermediate shallow state; the capture of an electron by the deep trap is spin dependent. 
To include magnetic field effects, Eq. (\ref{eq:Imax0}) must be modified as described next. 

The maximum recombination current is related to the recombination rate per area, $U_s$, by $I_{max} = q A U_s$.
The spin-dependent capture of a carrier electron by a deep trap is a two step process: (1) the carrier electron is first weakly localized by a shallow state in the vicinity of the defect. 
At low temperatures, neutral donors are a candidate for the shallow state. EDMR has resolved hyperfine structure in phosphorous-doped Si where recombination is mediated by charge transfer between $^{31}$P and the P$_{b0}$ defect.\cite{Morishita2009, Morishita2011, Dreher2015}  At room temperature this state is most likely an excited state of the defect which may be near the conduction band.\cite{Rong1991, Boehme2004, Friedrich2005, Hori2019} We assume this going forward.
(2) the electron in the now charged excited state of the deep paramagnetic trap reduces to its charged ground state if the singlet condition for the spin pair is met. We modify the theory of Shockley, Read\cite{Shockley1952},  and Hall\cite{Hall1952} to include the two-step capture (into the ground state) by the trap. The rate of capture of a conduction electron into the trap ground state is $c_n N_t$ where $c_n$ is the capture parameter (with dimension rate $\times$ volume) and  $N_t$ is the density of traps. If the two steps are each accomplished with respective capture parameters $c_{t^*}$ and $c_t$ then the 
total capture parameter of a conduction electron into the deep level trap by way of its excited state is
$c_{n}  = c_{t^*} c_t/(c_{t^*} + c_t) \approx  c_t$.
The last approximation is made since the transition rate to the trap ground state (large energy difference) is much smaller than the capture of a conduction electron by the excited state (small energy difference).
Unlike for holes, $c_n$ should not be expressed as $\sigma_n v_{th}$ because the electron is already situated near the defect (in the shallow state).
The calculation can then proceed  following that of Shockley, Read\cite{Shockley1952},  Hall\cite{Hall1952},  and Fitzgerald and Grove\cite{Fitzgerald1968}.
These calculations assume that traps have a constant density throughout gap but the recombination process is dominated by traps near the center. 
For a single trapping energy,
\begin{equation}
    U_s = n_t \frac{c_n c_p }{c_n (n_i + n_s) + c_p (n_i + p_s)} (n_s p_s - n_i^2), 
\end{equation}
where $n_{t}$ is the areal density of such centers and $n_s$, $p_s$ are electron and hole densities.
$U_s$ is maximized when the difference between $n_s$ and $p_s$ is minimized which occurs when $n_s = p_s = n_i e^{q |V_F|/2k_B T}$.
Under a typical forward bias, the  maximum recombination current approximates as
\begin{equation}\label{eq:maxCurrentSingle}
I_{max} = q n_t A \Sigma_1 n_i e^{q|V_F|/2k T}~\text{with } ~\Sigma_1 = \frac{c_n }{c_n/c_p+1}.
\end{equation}
More realistically, there is a distribution of trap energies within the band gap. Since recombination is dominated by traps near mid gap, the rate is largely immune to the shape of the distribution.\cite{Fitzgerald1968} For a uniform distribution, the recombination rate per unit area is:
\begin{equation}\label{eq:US1}
 U_s=\int_{E_v}^{E_c} \frac{c_{n} c_p  D_{st} (p_s n_s - n_i^2) ~dE_{st}}{c_{n}( n_s +  n_i e^{(E_{st} - E_i)/k_B T})  + c_p (p_s  +  n_i e^{-(E_{st} - E_i)/k_B T})} 
\end{equation}
\normalsize
where $E_{st}$ is the energy of the recombination center and $D_{st}$ is the areal density per energy of such centers, and $c_p = \sigma_p v_{th}$. The quantities $n_s $ and $p_s$ have dimensions of inverse volume. 
This calculation assumes the cross sections or capture parameters are independent of energy. 

This integral in Eq.~(\ref{eq:US1}) can be determined to be approximately\footnote{The intrinsic Fermi level, $E_i$, is assumed to be at mid gap such that $(E_v - E_i )/ k_B T \ll 0$ and $(E_c - E_i )/ k_B T \gg 0$. Also the forward bias should not be large enough to be comparable to $(E_c - E_v)/e$. }
\begin{equation}
U_s = \sqrt{c_n c_p}  k_B T D_{st} \frac{\text{ln}(x + \sqrt{x^2 - 1})}{n_i\sqrt{x^2 - 1}}(p_s n_s - n_i^2)
\end{equation}
with $
x = \frac{p_s}{2 n_i}\sqrt{\frac{\sigma_p v_{th}}{c_n}}  + \frac{n_s}{2 n_i}\sqrt{\frac{c_n}{\sigma_p v_{th}}}$,
where spin centers in a band of width $\frac{1}{2} q V_F \gg kT$ centered at mid gap contribute to the recombination rate. As a consequence the single-level approximation becomes more accurate at low temperatures. 
The function $\text{ln}(x + \sqrt{x^2 - 1})/\sqrt{x^2 - 1}$ decays monotonically from a maximum value of $\pi/2$. Thus maximizing the recombination rate entails minimizing the quantity $x$. Given that $n_i$ is the intrinsic carrier density, the only way to do so is by minimizing the difference between $n_s$ and $p_s$. 
Just as for Fitzgerald and Grove, the minimum values for these two are $n_i e^{q |V_F|/2k_B T}$ which can be reached by tuning the gate voltage. Note that the unequal cross sections do not change this criterion.

\begin{figure}[ptbh]
 \begin{centering}
        \includegraphics[scale = 0.2,trim = 0 0 920 0, angle = -0,clip]{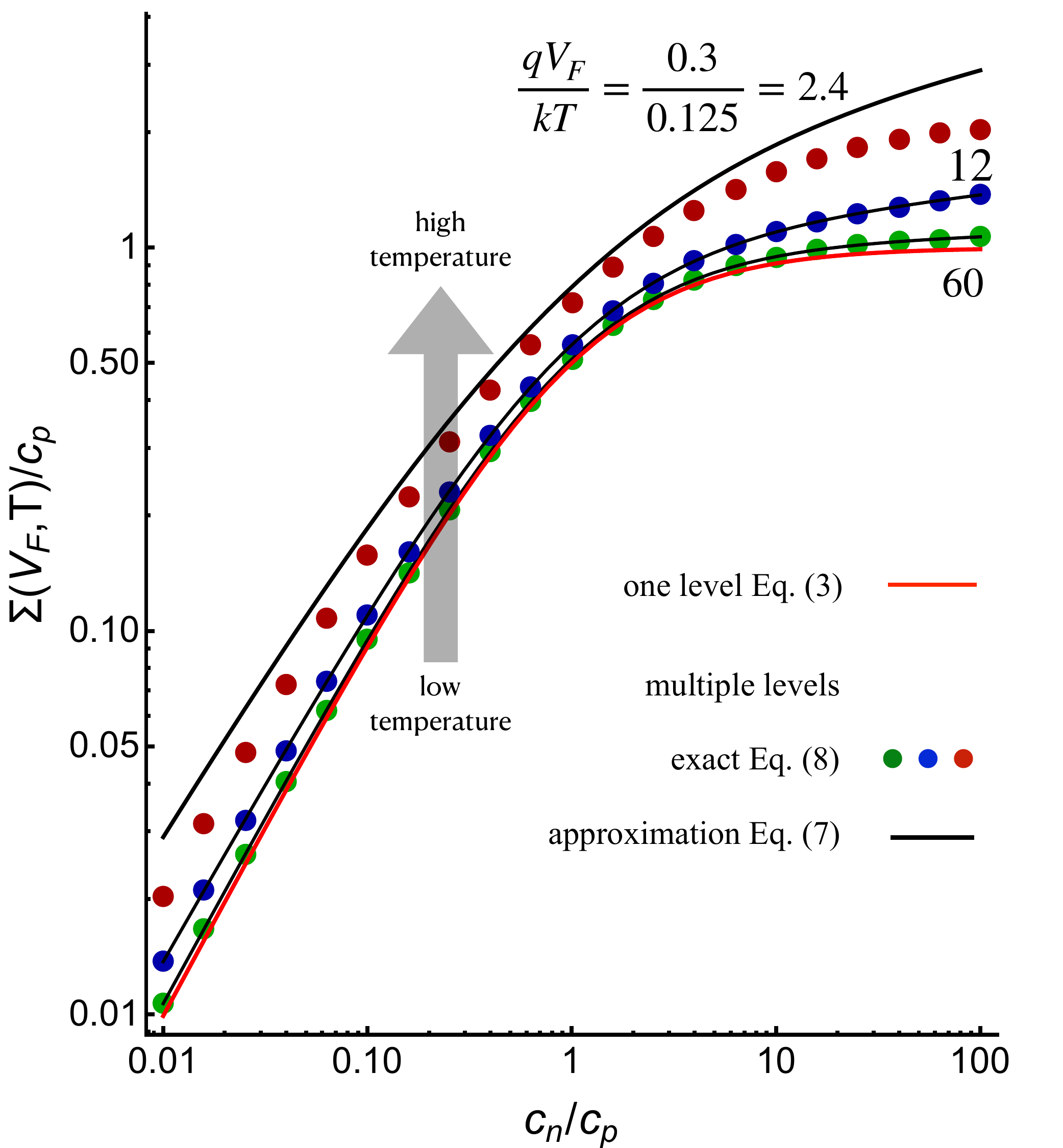}
        \caption[]
{$I_{max}$, evaluated at different $qV_F/kT$, for three different approximations: one level using $\Sigma_1$, uniform distribution (exact using Eq.~\ref{eq:exactSigma}) and approximate using Eq.~(\ref{eq:approxSigma}). The approximation performs poorly at higher temperature as expected.   
}
\label{fig:Imax-comps} 
        \end{centering}
 \end{figure}

Substituting in these constraints on the electron and hole densities yields, in the limit of large $|V_F|/2kT$,
\begin{equation}
U_s\approx  D_{st} n_i q |V_F| e^{q |V_F|/2k_B T} \frac{c_n c_p}{c_n + c_p} \left[1 +  \frac{\text{ln}(\sqrt{\frac{c_p}{c_n}}  + \sqrt{\frac{c_n}{c_p}} )}{\frac{q |V_F|}{2k_B T}}\right]  .
\end{equation}
where 
\begin{equation}\label{eq:approxSigma}
    \Sigma_a/c_p = \frac{c_n}{c_n + c_p} \left[1 +  \frac{\text{ln}(\sqrt{\frac{c_p}{c_n}}  + \sqrt{\frac{c_n}{c_p}} )}{\frac{q |V_F|}{2k_B T}}\right]
\end{equation}
is defined as the electron capture efficiency. A similar quantity is defined for the exact calculation of Eq. (\ref{eq:US1}):
\begin{equation}\label{eq:exactSigma}
    \Sigma_e = \frac{U_s}{n_i D_{st} q|V_F|e^{q|V_F|/2 kT}}.
\end{equation}
In the case of equal capture coefficients, the result of Eq. (\ref{eq:Imax0}) is retained save for a $\text{ln}(2)$ correction which is safely ignored since $q |V_F|/2k_B T \gg 1$.
However the $\text{ln}$ term can be sizable if the electron and hole capture are unbalanced which is usually the case. 
Figure \ref{fig:Imax-comps} presents the electron capture efficiencies for the exact calculations of Eq.~\ref{eq:exactSigma}, and the approximate calculations of eqs. \ref{eq:maxCurrentSingle} and \ref{eq:approxSigma}. 
At the highest temperature, the approximate solution strays from the exact as the condition of $q |V_F|/2k_B T \gg 1$ is violated.
For small $c_n$ compared to $\sigma_p v_{th}$, the dependence is linear.  We consider it in this way: electron-hole recombination occurs rapidly once the defect site becomes negatively charged which makes electron capture the limiting step toward quicker recombination. 
A defect like P$_{b0}$ at the Si/SiO$_2$ interface is in the regime $c_n \ll c_p=\sigma_p v_{th}$ since the trap will be negatively charged when capturing a hole but neutral when capturing an electron. 

The spin dependence enters the recombination through the electron capture coefficient, $c_n \approx c_t$. Of the electrons that enter the shallow state, $k_S \rho_S/g$ of those electrons will be successfully captured by the deep level.
If $c_{n,0}$ was the coefficient if all shallow electrons were to be captured, the spin dependence of the process modifies the coefficient to be $k_S \rho_S c_{n,0}/g$. $\rho_s$ is the steady state probability that a singlet spin pair occupies the two sites and is found from the stochastic Liouville equation which is demonstrated next.

To determine $\rho_s$, we solve for the spin-density matrix $\rho$ which fully accounts for the evolution of the spin-pair undergoing any number of spin interactions. This evolution  is governed by the stochatic Liouville equation:
\begin{equation}\label{eq:sle}
\frac{\partial \tilde{\rho}}{\partial t} = -\frac{i}{\hbar} [\tilde{H}, \tilde{\rho}] - \frac{k_S+k_D}{2} \{ P_S, \tilde{\rho} \}  - \frac{k_D}{2} \{ P_T, \tilde{\rho} \}+  \frac{g}{\text{Tr}\mathbb{1}}\mathbb{1}
\end{equation}
where $\tilde{\rho}$ is the density matrix in the rotating reference frame
and
$
\tilde{H} = g_e \mu_B  (B_0 + \frac{\hbar \Omega}{g_e \mu_B }) (S_{z,1} + S_{z,2}) + \hbar \Omega I_{z} + 
 g_e \mu_B  \bm{I}\cdot \hat{\mathbf{A}} \cdot \bm{S}_1 +  g_e \mu_B B_1 (S_{x,1} + S_{x,2})
$
where we assume the hyperfine tensor ($\hat{\mathbf{A}}$) between defect spin ($\bm{S}_1$) and nuclear spin ($\bm{I}$) possesses axial symmetry.
The first two terms are the static Zeeman interaction for the electrons and nuclei ($\Omega$ is the microwave frequency). The third term is the hyperfine interaction, and the fourth term is the interaction of the spins with the transverse oscillating field. 
In Eq.~(\ref{eq:sle}), the first term on the right-hand side is the Liouville equation for the density matrix, describing the coherent evolution of the density matrix.  The second and third terms signify the random processes of spin capture and spin dissociation of the spin pairs which, in general, may depend on their spin configuration (singlet or triplet combination occur at rates $k_S$ and $k_T$, respectively). Assuming small spin-orbit interactions, we take $k_T = 0$; triplets are not captured by the deep defect. The rate of singlet capture depends on the occupation of states so is written as $k_S \text{Tr}[P_S \rho(t)] = k_S \rho_S(t)$. A rate $k_D$ describes the dissociation of the spin pair. $P_S$ is the singlet projector.

\begin{figure}[ptbh]
 \begin{centering}
        \includegraphics[scale = 0.2,trim = 5 5 550 90, angle = -0,clip]{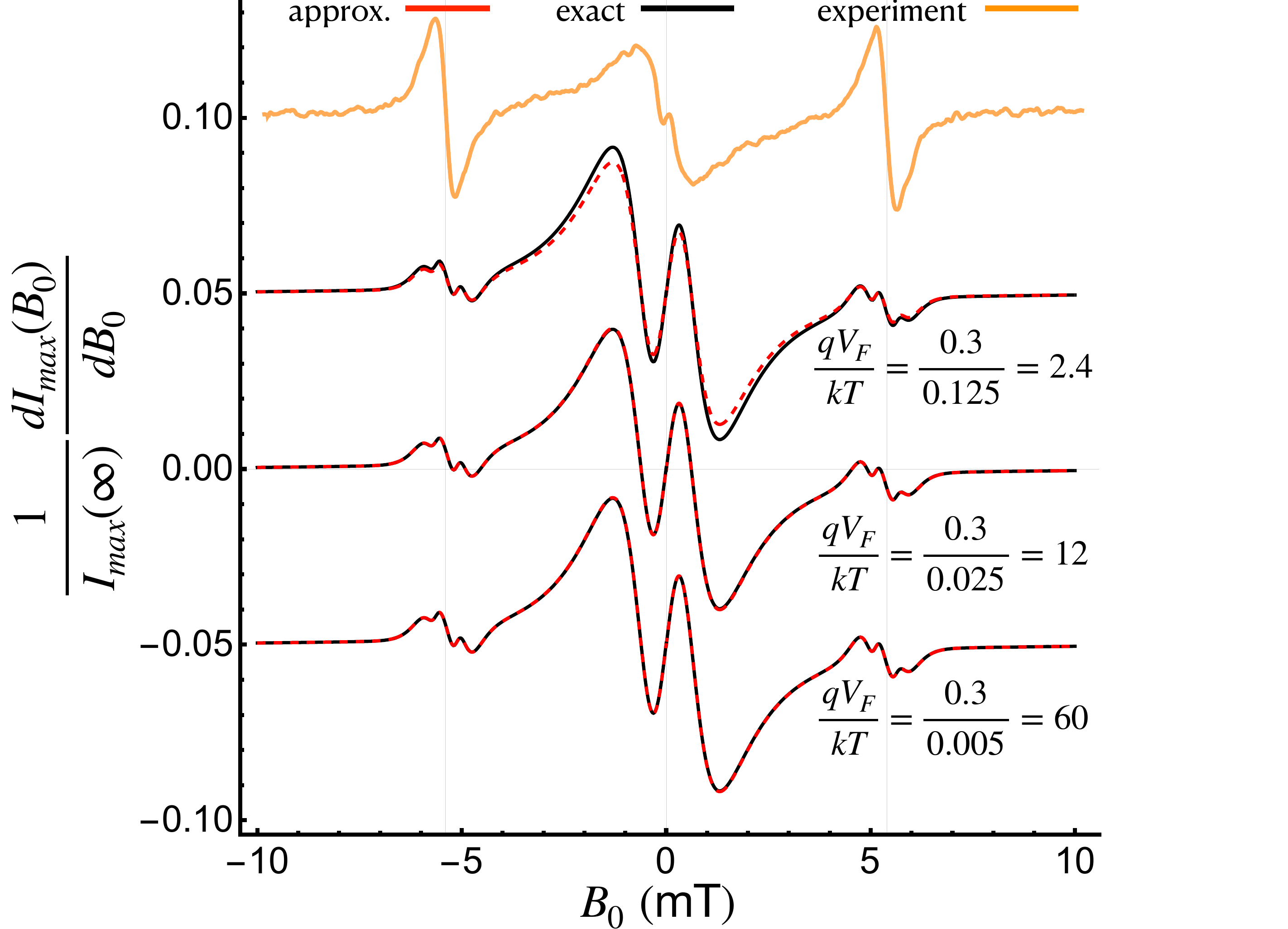}
        \caption[]
{Orange line is experimental trace reported in Ref. \onlinecite{Ashton2019}. Red and black lines are normalized current derivatives at three different temperatures using $k_S = 0.16$ ns$^{-1}$, $k_D = 0.08$ ns$^{-1}$, $B_1 = 0.2$ mT, $A = 1$ mT, and $c_{n,0}/c_p = 0.01$. Approximate result is Eq.~(\ref{eq:maxCurrentFinal}). Thin vertical lines denote the resonances due to a 151 MHz field. All curves are offset for clarity.
}\label{fig:MR-comps} 
        \end{centering}
 \end{figure}

Our final approximate expression for the maximum recombination current is
\begin{equation}\label{eq:maxCurrentFinal}
I_{max}  = \Sigma(B_0) q^2 A D_{st} n_i |V_F| e^{q |V_F|/2 k_B T},
\end{equation}
where 
\begin{equation}
    \Sigma \approx \frac{k_S}{g}\rho_S(B_0) c_{n,0} \left[1 +  \frac{\text{ln}(\sqrt{\frac{g}{k_S}\frac{1}{\rho_S(B_0)} \frac{c_p}{c_{n,0}}}  + \sqrt{\frac{k_S}{g}\rho_S(B_0)\frac{c_{n,0}}{c_p}} )}{\frac{q |V_F|}{2k_B T}}\right].
\end{equation}
This expression is valid as long as the bias condition for maximum current is unchanged by the magnetic field. The optimum current's dependence on bias has not been systematically studied in relation to magnetic field but one study shows very little change in the bias condition. \cite{Campbell2007} 
All other quantities in $I_{max}$ are considered to be spin or magnetic field-independent.
To maximize sensitivity, EDMR and NZFMR experiments are often carried out using lock-in amplification which manifests a line shape that appears as the derivative of a conventional response observed under a traditionally used magnetic field modulation scheme. Such a response is shown in Fig. \ref{fig:MR-comps}. To compare to our calculation $I^{-1}_{max}(\infty) dI_{max}(B_0)/dB_0$ is plotted for three different ratios of bias and temperature. 
Our intent here is to demonstrate general agreement of resonant and near-zero field responses between computation and experiment. Fitting the data is beyond the scope of this Letter.
When investigating the saturation behavior of EDMR on $B_1$, we do find a sign change (EDMR amplitude inverts) which has been attributed to the spin-Dicke effect. \cite{Waters2015, Mkhitaryan2018}
In the low $B_1$ regime, our results demonstrate a quadratic in $B_1$ relationship in agreement which has been observed in EDMR experiments.\cite{Kawachi1997, Eickelkamp1998}

The goal of this article has been to provide a sound theoretical basis for the near-zero-field magnetoresistance phenomena present at oxide-semiconductor interfaces in technologically relevant devices. While here the focus as been on spin-dependent recombination in MOSFETs, our approach applies more broadly. Two examples are (1) spin-dependent recombination in phosphorous-doped Si where shallow donors mediate recombination through deep defects\cite{Morishita2009, Morishita2011, Dreher2015}  and (2) spin-dependent trap-assisted transport where NZFMR is also observed.\cite{Frantz2020}

\textbf{Supplementary Material. --} 
The supplemental material expands on the solutions to Eq. \ref{eq:sle}, derives the Hamiltonians with the rotating wave approximation, and elaborates on the experimental details for the experimental trace of Fig. \ref{fig:MR-comps}.

\textbf{Acknowledgements. --} 
The project or effort depicted was or is sponsored by the Department of the Defense, Defense Threat Reduction Agency under Grants HDTRA1-18-0012 and HDTRA1-16-0008. 
This material is based upon work supported by the Air Force Office of Scientific Research under award number FA9550-22-1-0308. 
The content of the information does not necessarily reflect the position or the policy of the federal government, and no official endorsement should be inferred. 

\textbf{Data Availability. --}
The data that support the findings of this study are available from the corresponding author upon reasonable request. 


%


\end{document}